\documentclass[aps,10pt,prd,twocolumn,preprintnumbers,amsmath,amssymb,nofootinbib,superscriptaddress,a4paper,showpacs]{revtex4-1}
\pdfoutput=1
\usepackage[T1]{fontenc}

\usepackage[english]{babel}
\usepackage[utf8]{inputenc}
\usepackage{graphicx}
\usepackage{color}
\usepackage{amsfonts,amsthm}
\usepackage{bm,bbm}
\usepackage{xcolor}
\usepackage{mathrsfs}
\usepackage{amsmath}

\newcommand{\lam}{\ensuremath{\lambda}}

\newcommand{\hf}{\frac{1}{2}}

\newcommand{\cO}{{\cal O}}
\newcommand{\cA}{{\cal A}}
\newcommand{\cAb}{{\overline{\cal A}}}
\newcommand{\cF}{{\cal F}}
\newcommand{\cFb}{{\overline{\cal F}}}
\newcommand{\cD}{{\cal D}}
\newcommand{\cDb}{{\overline{\cal D}}}
\newcommand{\cQ}{{\cal Q}}
\newcommand{\cU}{{\cal U}}
\newcommand{\cN}{{\cal N}}
\newcommand{\cUb}{{\overline{\cal U}}}
\newcommand{\Tr}{{\rm Tr\;}}

\newcommand{\vn}{ {\bf n} }

\newcommand{\Ibb}{\ensuremath{\mathbb I} }
\newcommand{\hatbmu}{\widehat{\boldsymbol {\mu}}}

\newcommand{\SB}{{\bar{S}}}
\newcommand{\CHI}{\ensuremath{\mbox{$\chi^2/\text{d.o.f.}$}}}

\def\nn{\nonumber}
\def\bec{\begin{center}}
\def\eec{\end{center}}
\def\beq{\begin{equation}}
\def\eeq{\end{equation}}
\def\bea{\begin{eqnarray}}
\def\eea{\end{eqnarray}}

\newcommand{\RT}{{r_{\tau}}}
\newcommand{\vac}{\mathcal{E}_{\mathrm{VAC}}}

\newcommand{\vev}[1]{\ensuremath{\left\langle #1 \right\rangle} }

\usepackage[colorlinks = true, linkcolor = blue,urlcolor  = blue, citecolor = blue, anchorcolor = blue]{hyperref}

\begin{document}

\title{Nonperturbative study of dynamical SUSY breaking in $\cN = (2, 2)$ Yang-Mills}

\author{Simon Catterall}
\email{smcatter@syr.edu}
\affiliation{Department of Physics, Syracuse University, Syracuse, NY 13244, USA}

\author{Raghav G. Jha}
\email{rgjha@syr.edu}
\affiliation{Department of Physics, Syracuse University, Syracuse, NY 13244, USA}

\author{Anosh Joseph}
\email{anosh.joseph@icts.res.in}
\affiliation{International Centre for Theoretical Sciences (ICTS-TIFR), \\ Tata Institute of Fundamental Research, \\ Bangalore, 560089 INDIA}

%%%%%%%%%%%%%%%%%%%

\begin{abstract}
We examine the possibility of dynamical supersymmetry breaking in two-dimensional ${\cal N} = (2, 2)$ supersymmetric Yang-Mills theory. The theory is discretized on a Euclidean spacetime lattice using a supersymmetric lattice action. We compute the vacuum energy of the theory at finite temperature and take the zero temperature limit. Supersymmetry will be spontaneously broken in this theory if the measured ground state energy is non-zero. By performing simulations on a range of lattices up to $96 \times 96$ we are
able to perform a careful extrapolation to the continuum limit for a wide range of temperatures. Subsequent extrapolations to the zero temperature limit yield an upper bound on the ground state energy density. We find the energy density to be statistically consistent with zero in agreement with the absence of dynamical supersymmetry breaking in this theory.
\end{abstract}

%%%%%%%%%%%%%%%%%%%%%%%%%%%%%%%%%%%%%%%%%%%%%%%%%%%%%%%%%%%

\maketitle

%%%%%%%%%%%%%%%%%%%%%%%%%%%%%%%%%%%%%%%%%%%%%%%%%%%%%%%%%%%
\section{Introduction}
\label{sec:intro}
%%%%%%%%%%%%%%%%%%%%%%%%%%%%%%%%%%%%%%%%%%%%%%%%%%%%%%%%%%%

The investigations of supersymmetric gauge theories on a spacetime lattice are important for understanding the non-perturbative structure of such theories and in particular they can address the question of whether dynamical supersymmetry breaking takes place in such theories. This is a crucial question for efforts to construct supersymmetric theories which go beyond the Standard Model since the low energy world is clearly not supersymmetric while non-renormalization theorems typically ensure that supersymmetry cannot break in perturbation theory \cite{Grisaru:1979wc}.

Unfortunately, there are a plethora of problems to overcome for lattice formulations of supersymmetric theories. Supersymmetry is a spacetime symmetry, which is generically broken by the lattice regularization procedure. Hence, the effective action of the lattice theory typically contains relevant supersymmetry breaking interactions. To achieve a supersymmetric continuum limit it is necessary to  fine tune the lattice couplings to these terms as the lattice spacing is reduced. Since generically there are very many such terms this is in practice impossible.  Some exceptions to this are - $\cN = 1$ super Yang-Mills where only a single coupling, the gluino mass, must be tuned. In addition, it has also been shown that fine-tuning to a supersymmetric
continuum limit is also possible for $\cN = (2,2) $ in two dimensions. Using
Wilson fermions, the only relevant parameter that has to be fine-tuned
is the scalar mass since the bare gluino mass is an irrelevant parameter. The
continuum value for the critical scalar mass is known up to one-loop
order in lattice perturbation theory and that has already been employed in the numerical simulations. See Ref. \cite{Montvay:2001aj, Suzuki:2005dx,August:2018esp} for discussions and references therein. 

The attempt to formulate supersymmetric theories on the lattice has a long history starting in Refs. \cite{Elitzur:1982vh, Banks:1982ut, Sakai:1983dg, Kostelecky:1983qu, Aratyn:1984bc, Scott:1983ha}. Recent approaches to this problem have focused on preserving a subalgebra of the full supersymmetry algebra which can protect the theory from some of these dangerous supersymmetry violating terms - for a review, see Ref. \cite{Catterall:2009it}. For supersymmetric theories with extended supersymmetry various supersymmetric lattice formulations exist. One approach that was pioneered by Cohen, Kaplan, Katz and \"Unsal in Refs. \cite{Cohen:2003xe, Cohen:2003qw, Kaplan:2005ta} is based on orbifolding and deconstruction of a supersymmetric matrix model. A second approach uses the idea of topological twisting to isolate appropriate nilpotent scalar supersymmetries that can be transferred to the lattice. Two independent discretization schemes have been proposed in this approach - that proposed by Sugino in Refs. \cite{Sugino:2003yb, Sugino:2004qd} where the fermions are associated with sites and  a geometrical approach in which fermions are generically associated with links \cite{Catterall:2003wd}\footnote{Yet another construction was formulated by D'Adda, Kanamori, Kawamoto and Nagata, \cite{DAdda:2005rcd} but was later shown to be equivalent to the orbifolding constructions when restricted to a sector containing a scalar supercharge \cite{Damgaard:2007eh}.}. In four spacetime dimensions, the geometrical approach has been used to construct  a supersymmetric lattice action for $\cN = 4$ SYM  \cite{Catterall:2005fd, Catterall:2014vka} and has been shown to be identical to the orbifolding constructions in Ref. \cite{Unsal:2006qp,Catterall:2007kn}. For an elaborate discussion on the relation between all these constructions, see Ref. \cite{Takimi:2007nn}. 

In this paper, we will study $\cN = (2, 2)$ super Yang-Mills (SYM) theory using the geometrical discretization scheme. It is the simplest two-dimensional supersymmetric theory that can be studied on the lattice. This theory is a particularly interesting theory in the continuum because of its exotic phases as discussed by Witten in Ref. \cite{Witten:1993yc}. This theory is conjectured to flow in the infrared (IR) to a conformal field theory. For recent developments, see Ref. \cite{Park:2016dpb}. The goal  of this paper is to calculate the vacuum energy density accurately for this theory and hence determine whether supersymmetry breaking occurs. It is well known \cite{Witten:1981nf} that the vacuum energy can be thought of as an order parameter for  SUSY breaking. The spontaneous breaking of supersymmetry in this two-dimensional theory has been considered theoretically in Ref. \cite{Hori:2006dk} and numerically in Refs. \cite{Kanamori:2009dk, Kanamori:2007yx}. In \cite{Hori:2006dk} it was conjectured that in fact supersymmetry may break in this theory. Related work for $\cN = (2, 2)$ super QCD on the lattice was described in \cite{Catterall:2015tta}. In the context of orbifold lattice theories, it was shown in Ref. \cite{Matsuura:2007ec} that the vacuum energy of these theories does not receive any quantum corrections in perturbation theory leaving only non-perturbative mechanisms to drive supersymmetry breaking.

In this four supercharge theory, unlike the sixteen supercharge case in two dimensions, the thermal instabilities at low temperatures are less severe and we can access relatively small temperatures without truncating the $U(1)$ degree of freedom as done in our recent work \cite{Catterall:2017lub, Jha:2017zad}. However, we have to use a small mass term to control the classical flat directions associated with the scalars. This small mass term was also implemented while exploring the phase structure at large $N$ using Sugino's lattice construction in Ref. \cite{Hanada:2009hq}. 

The plan of this paper will be as follows. In Sec. \ref{sec:theory} we review the lattice construction for $\cN = (2, 2)$ SYM on a two-dimensional square lattice. Then in Sec. \ref{sec:computation} we mention results on the phase of the pfaffian, discuss our procedure of extracting the ground state energy and comment on the $\cO(a)$ improved action we use for the analysis. We end the paper with conclusions and brief discussion in Sec. \ref{sec:conclude}. 

%%%%%%%%%%%%%%%%%%%%%%%%%%%%%%%%%%%%%%%%%%%%%%%%%%%%%%%%%%%
\section{Two-dimensional $\cN = (2, 2)$ Lattice SYM}
\label{sec:theory}
%%%%%%%%%%%%%%%%%%%%%%%%%%%%%%%%%%%%%%%%%%%%%%%%%%%%%%%%%%%

The two-dimensional $\cN = (2, 2) $ SYM theory is the simplest supersymmetric gauge theory which admits topological twisting \cite{Witten:1988ze} and thus satisfies the requirements for a supersymmetric lattice construction following the prescription given in Refs. \cite{Catterall:2004np, Catterall:2006jw}, where the first numerical simulations of this construction were performed. The theory has global symmetry group $G = SO(2)_E \times SO(2)_{R_1} \times U(1)_{R_2}$, where $SO(2)_E$ is the two-dimensional Euclidean Lorentz rotation symmetry, $SO(2)_{R_1}$ is the symmetry due to reduced directions and $U(1)_{R_2}$ is the R-symmetry of the parent four-dimensional $\cN =1$ SYM theory. This theory can be twisted in two inequivalent ways (the A-model and B-model twists) depending on how we embed $SO(2)_E$ group into $SO(2)_{R_1} \times SO(2)_{R_2}$ the internal symmetry group. 

We are interested in the B-model twist, which gives rise to a strictly nilpotent twisted supersymmetry charge. After twisting, the fields and supersymmetries are expressed as representations of the the twisted Euclidean Lorentz group
\beq
SO(2)' = {\rm diag} \Big( SO(2)_E \times SO(2)_{R_1} \Big).
\eeq 
The action of continuum $\cN = (2, 2)$ SYM takes the following $\cQ$-exact form after twisting
\beq
S = \frac{N}{2 \lam} \cQ \int d^2x\,\Psi,
\label{2daction_twisted}
\eeq
where 
\beq
\Psi = \Tr \left( \chi_{ab} \cF_{ab} + \eta [ \cDb_a, \cD_b ] - \hf \eta d \right),
\eeq
and $\lam = g^2 N$ is the 't Hooft coupling. We use an anti-hermitian basis for the generators of the gauge group with $\Tr (T_a T_b)  = - \delta_{ab}$.

\begin{figure}[htb]
\begin{center}
\includegraphics[width=0.45\textwidth]{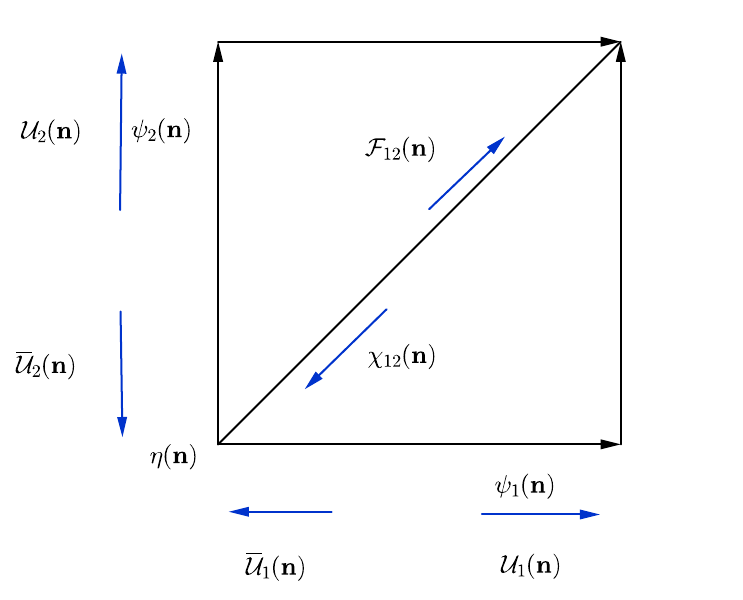}
\end{center}
\caption{\label{fig:lattice-unit-cell}The unit cell and field orientations of the two-dimensional $\cN = (2, 2)$ lattice SYM theory.}
\end{figure}

The four degrees of freedom appearing in the above action are just the twisted fermions $(\eta, \psi_a, \chi_{ab})$ and a complexified gauge field $\cA_a$. The complexified field is constructed from the usual gauge field $A_a$ and the two scalars $B_a$ present in the untwisted theory: $\cA_a = A_a + iB_a$.
The twisted theory is naturally written in terms of the complexified covariant derivatives
\beq
\cD_a = \partial_a + \cA_a, \quad \quad \cDb_a = \partial_a + \cAb_a,
\eeq
and complexified field strengths
\beq
\cF_{ab} = [\cD_a, \cD_b], \quad \quad \cFb_{ab} = [\cDb_a, \cDb_b].
\eeq
The nilpotent supersymmetry transformations associated with the scalar supercharge $\cQ$ are given by
\bea
\cQ\; \cA_a &=& \psi_a, \nn \\
\cQ\; \psi_a &=& 0, \nn \\
\cQ\; \cAb_a &=& 0, \nn \\
\cQ\; \chi_{ab} &=& -\cFb_{ab}, \nn \\
\cQ\; \eta &=& d, \nn \\
\cQ\; d &=& 0.
\eea
Performing the $\cQ$-variation on $\Psi$ and integrating out the auxiliary field $d$ yields
\bea
S &=& \frac{N}{2\lam} \int \Tr \bigg( - \cFb_{ab} \cF_{ab} + \hf [ \cDb_a, \cD_a ]^2 \nonumber \\ 
&-&\chi_{ab} \cD_{\left[a\right.} \psi_{\left.b\right]} - \eta \cDb_a\psi_a \bigg). 
\label{2d-twist_action}
\eea
The prescription for discretization is straightforward. The complexified gauge fields are mapped to complexified Wilson links
\beq
\cA_a(x) \rightarrow \cU_a(\vn),
\eeq
living on the links of a square lattice with integer-valued basis vectors along two directions, 
\beq
\hatbmu_1 = (1, 0), \quad \quad \hatbmu_2 = (0, 1).
\eeq
They transform in the appropriate way under $U(N)$ lattice gauge transformations
\beq
\cU_a(\vn) \to G(\vn) \cU_a(\vn) G^\dagger(\vn+\hatbmu_a).
\eeq
Supersymmetry invariance then implies that $\psi_a(\vn)$ live on the same links and transform identically. The scalar fermion $\eta(\vn)$ is associated with a site and transforms the following way under gauge transformations
\beq
\eta(\vn) \to G(\vn) \eta(\vn) G^\dagger(\vn).
\eeq
The field $\chi_{ab}(\vn)$, as a 2-form, should be associated with a plaquette. In practice, we introduce diagonal links running through the center of the plaquette and choose $\chi_{ab}(\vn)$ to lie with opposite orientation along those diagonal links. This orientation ensures gauge invariance. Fig. (\ref{fig:lattice-unit-cell}) shows the unit cell of the lattice theory with field orientation assignments. 

%%%%%%%%%%%%%%%%%%%%%%%%%%%%%%%%%%%%%
%%%%%%%%%%%%%%%%%%%%%%%%%%%%%%%%%%%%%

The continuum covariant derivatives are replaced by covariant difference operators and they act on the twisted fields the following way 
\bea
\cDb^{(-)}_a f_a(\vn) &=& f_a(\vn) \cUb_a(\vn) - \cUb_a(\vn - \hatbmu_a) f_a(\vn - \hatbmu_a), \nonumber  \\ 
\cD^{(+)}_a f_b(\vn) &=& \cU_a(\vn) f_b(\vn + \hatbmu_a) - f_b(\vn) \cU_a(\vn + \hatbmu_b). \nonumber 
\eea
The lattice field strength is given by $\cF_{ab}(\vn) = \cD^{(+)}_a \cU_b(\vn)$, and is anti-symmetric. It transforms like a lattice 2-form and yields a gauge invariant loop on the lattice when contracted with $\chi_{ab}(\vn)$. Similarly, the term involving the covariant backward difference operator, $\cDb^{(-)}_a \cU_a(\vn)$, transforms as a 0-form or site field and hence can be contracted with the site field $\eta(\vn)$ to yield a gauge invariant expression.

The lattice action is $\cQ$-exact
\bea
S &=& \frac{N}{2 \lam} \sum_{\vn} \Tr \cQ \Big( \chi_{ab}(\vn) \cD_a^{(+)} \cU_b(\vn) \nonumber \\  
&+& \eta(\vn) \cDb_a^{(-)}\cU_a(\vn) - \hf \eta(\vn) d(\vn) \Big).
\eea
Applying the $\cQ$ transformation on the lattice fields and integrating out the auxiliary field $d$, we obtain the gauge invariant and $\cQ$ supersymmetric lattice action
\bea
\label{eq:2d-latticeaction}
S &=& S_B + S_F,
\eea
where the bosonic action is
\bea
S_B &=& \frac{N}{2 \lam} \sum_{\vn} \Tr \Big( \cF_{ab}^\dagger(\vn) \cF_{ab}(\vn) + \hf \Big(\cDb_a^{(-)} \cU_a(\vn)\Big)^2\Big), \nonumber 
\eea
and the fermionic piece
\bea
S_F &=& \frac{N}{2 \lam} \sum_{\vn} \Tr \Big( - \chi_{ab}(\vn) \cD^{(+)}_{[a} \psi_{b]}(\vn) - \eta(\vn) \cDb^{(-)}_a \psi_a(\vn) \Big). \nonumber 
\eea
It was correctly noted in Ref. \cite{Kanamori:2008bk} that for simulation purposes, we need to add a small supersymmetry breaking scalar potential to stabilize the $SU(N)$ flat directions of the theory. We add a single trace deformation term to the action in Eq. (\ref{eq:2d-latticeaction}) as, 
\beq
\label{eq:single_trace}
S_{\rm soft} = \frac{N}{2 \lam} \mu^2 \sum_{\vn,\ a} \Tr \bigg( \cUb_a(\vn) \cU_a(\vn) - \Ibb_N \bigg)^2,
\eeq
with a tunable parameter $\mu$. Exact supersymmetry at $\mu = 0$ ensures that all $\cQ$-breaking terms vanish as some (positive) power of $\mu$.

%%%%%%%%%%%%%%%%%%%%%%%%%%%%%%%%%%%%%%%%%%%%%%%%%%%%%%%%%%%
\section{Lattice Simulations}
\label{sec:computation}
%%%%%%%%%%%%%%%%%%%%%%%%%%%%%%%%%%%%%%%%%%%%%%%%%%%%%%%%%%%

We simulate the theory on a square lattice with anti-periodic boundary conditions (aPBC) for fermions in the temporal direction. The physical size of the lattice is $\beta \times L$, where $\beta$ is the dimensionful temporal extent and $L$ the dimensionful spatial extent. We denote the lattice spacing as $a$ while $N_t$ is the number of lattice sites along the temporal direction and $N_x$ number of sites along the spatial direction. Thus the dimensionful quantities are $\beta = a N_t$ and $L = a N_x$. In our case the lattice is symmetric: $N_t = N_x$. 

In two dimensions, the 't Hooft coupling $\lambda$ is dimensionful and we can construct the dimensionless temporal circle size, 
\beq
r_\tau = \sqrt{\lambda} \beta.
\eeq
The quantity $r_\tau$ also serves as the effective coupling. Its inverse is the dimensionless temperature $t$. Since we have only considered symmetric lattices, the spatial circle size is the same as the temporal circle size, $r_x = r_\tau$. As discussed above we use a small mass parameter $\mu = \zeta \frac{r_\tau}{N_t} = \zeta \sqrt{\lambda} a$ to regulate potential divergences associated with the flat directions. As for case of sixteen supercharge theory in two dimensions \cite{Catterall:2017lub, Jha:2017zad}, we extrapolate all our results to $\mu = 0$.

To examine the question of supersymmetry breaking we consider the system at non-zero temperature and subsequently take the temperature to zero {\it after} taking the limits $\zeta \to 0$ and $a \to 0$. A non-zero value of the vacuum energy would indicate supersymmetry breaking. Notice that if supersymmetry is intact in a finite volume, it is unbroken even in infinite volume \cite{Witten:1982df}.

We compute the ground state energy density in two-dimensional $\cN = (2, 2)$ SYM using the publicly available code presented in Ref. \cite{Schaich:2014pda}. In the four-supercharge case, the expression for the \emph{effective} bosonic action, which is related to the dimensionless energy density we measure, was first given in Ref. \cite{Catterall:2008dv}. 

We can have two different definitions for the ground state energy based on whether we take the massless (scalar mass) limit followed by the continuum limit or vice versa. In both cases, the zero temperature limit is taken at the end. Thus, we have 
\beq
\label{eq:energy1} 
\frac{\vac^{\prime}}{N^2 \lambda} = \lim_{\beta\to\infty} ~~\lim_{\mathrm{a} \to 0} ~~\lim_{\mu \to 0} \Bigg \langle \mathrm{VAC} \Bigg | \left( \frac{-2\SB}{N^2 \lam} \right) \Bigg | \mathrm{VAC}  \Bigg \rangle,
\eeq
and 
\beq
\label{eq:energy2} 
\frac{\vac}{N^2 \lambda} = \lim_{\beta\to\infty} ~~ \lim_{\mu \to 0} ~~\lim_{\mathrm{a} \to 0} \Bigg \langle \mathrm{VAC} \Bigg | \left( \frac{-2\SB}{N^2 \lam} \right) \Bigg | \mathrm{VAC}  \Bigg \rangle,
\eeq
where, 
\beq
\SB = \frac{1}{L \beta} \left( S_B - \frac{3}{2} N^2 N_x N_t \right).
\eeq 
%%%%%%%%%%%%%%%%%%%%%%%%%%%%%%%%%%%
\begin{figure}[htb]
\begin{center} 
\includegraphics[width=0.45\textwidth]{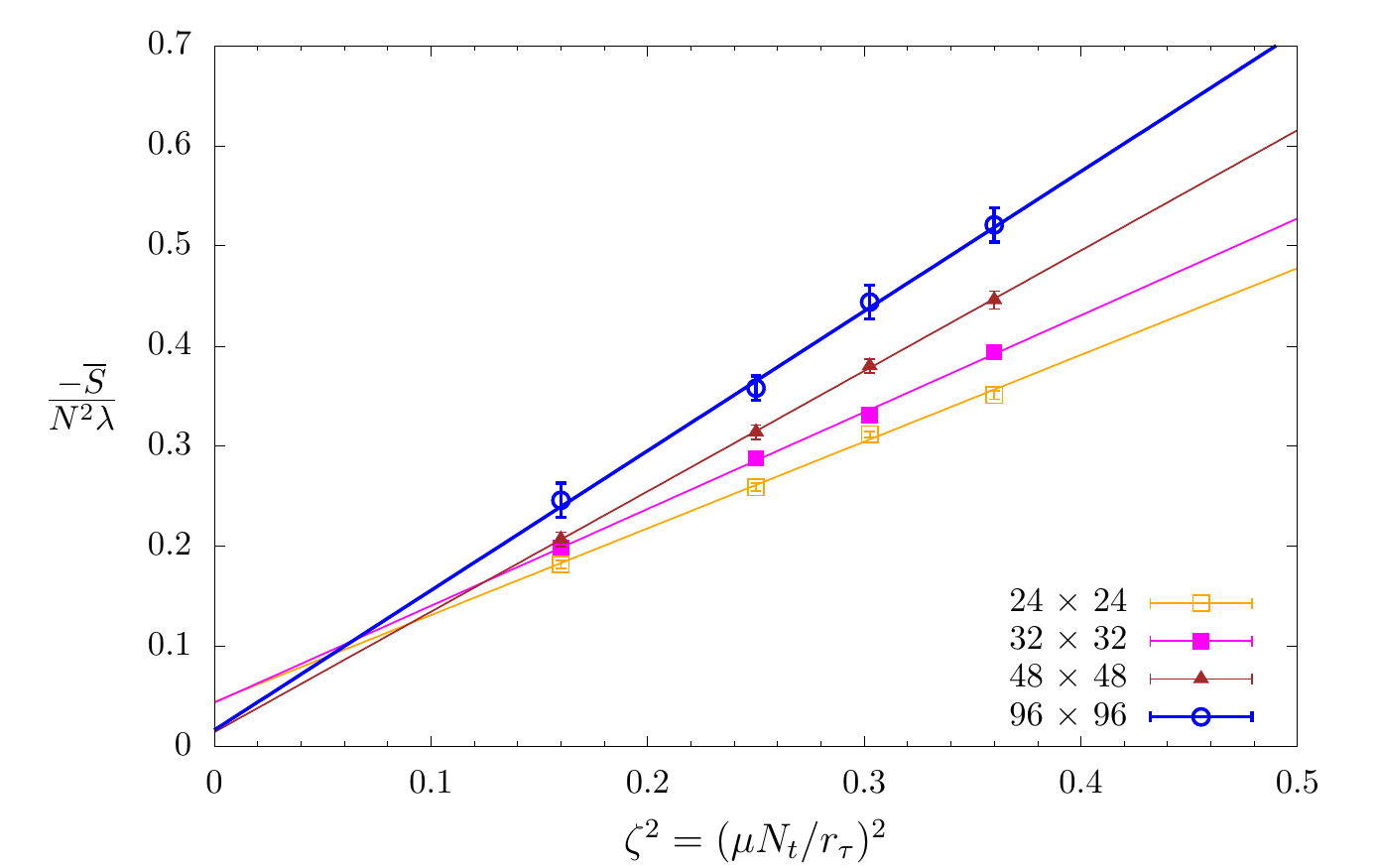}
\end{center}
\caption{\label{fig:mass_U3}The $\zeta^2 \to 0$ extrapolation of the ground state energy density for $U(3)$, $\RT = 9$.}
\end{figure}
%%%%%%%%%%%%%%%%%%%%%%%%%%%%%%%%%%%
%%%%%%%%%%%%%%%%%%%%%%%%%%%%%%%%%%%
\begin{figure}[htb]
\begin{center} 
\includegraphics[width=0.45\textwidth]{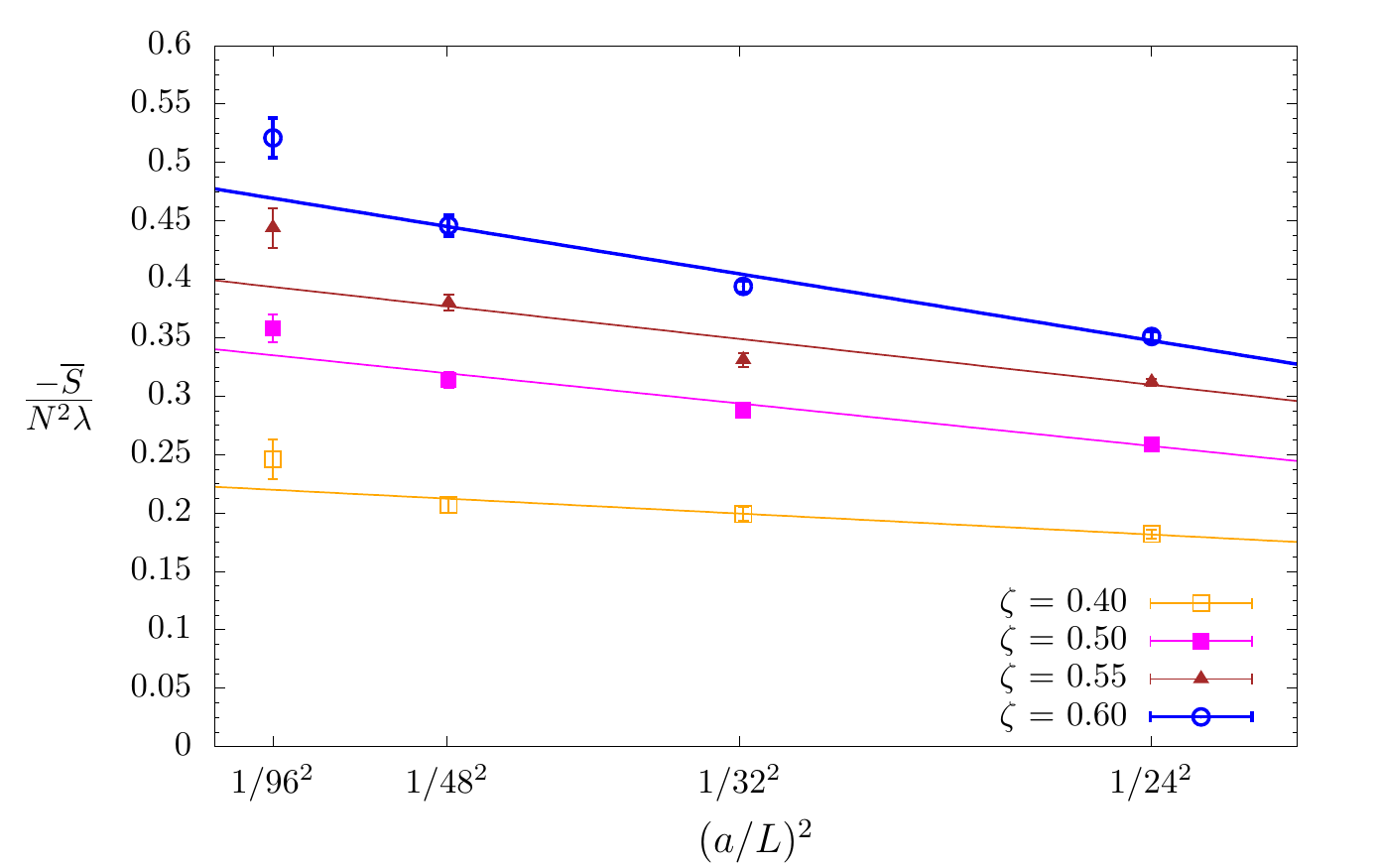}
\end{center}
\caption{\label{fig:cont_U3}The $(a/L)^2 \to 0$ extrapolation of the ground state energy density for $U(3)$, $\RT=9$.}
\end{figure}
%%%%%%%%%%%%%%%%%%%%%%%%%%%%%%%%%%%
%%%%%%%%%%%%%%%%%%%%%%%%%%%%%%%%%%%%%%%%%%%%%%%%%%%%%%%%%%%

We provide the simulation data in Tables (\ref{tab:actionU2}) and (\ref{tab:actionU3}). It is clear from the tables that the order of taking these \emph{different} limits is consistent within errors and we will quote results only for $\frac{\vac}{N^2 \lambda}$.
\begin{figure}[htb]
\begin{center} 
\includegraphics[width=0.45\textwidth]{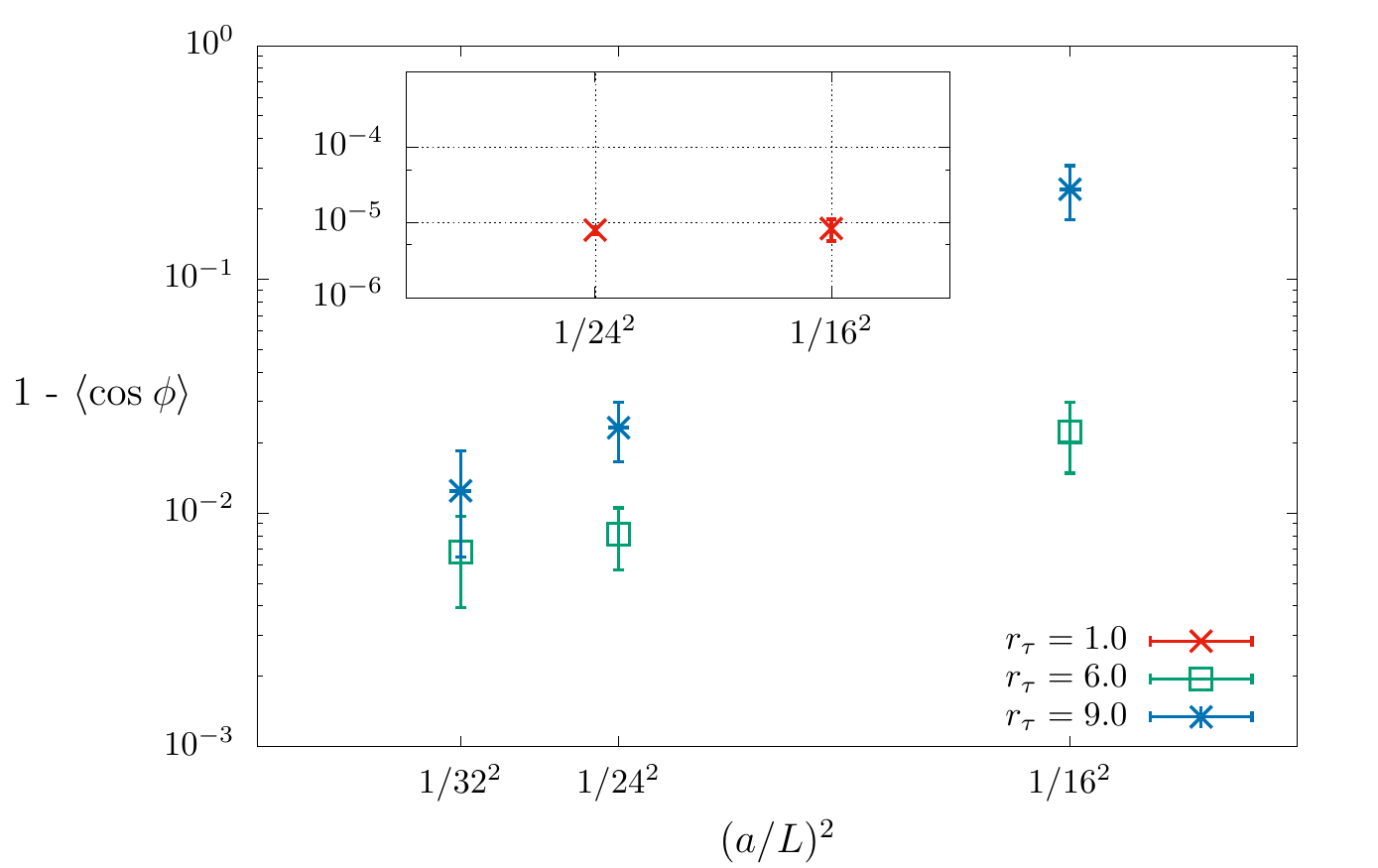}
\end{center}
\caption{\label{fig:pfaffian1}Pfaffian phase fluctuations, $1 - \vev{\cos\phi}$, for some U(3) ensembles used in this work. We have measured the phase for three couplings used in this work. We keep the mass parameter, $\zeta = 0.50$ for all couplings. Note that at sufficiently weak couplings, large lattices are not needed to control sign problem.}
\end{figure}
%%%%%%%%%%%%%%%%%%%%%%%%%%%%%%%%%%%
%%%%%%%%%%%%%%%%%%%%%%%%%%%%%%%%%%%
We integrate out the fermions to produce a Pfaffian, which in turn is represented by square root of a determinant. The fermion determinant with a fractional power can be simulated using Rational Hybrid Monte Carlo (RHMC) algorithm \cite{Clark:2004cp}. In the simulations we used the absolute value of the Pfaffian. The phase of the Pfaffian may be incorporated back in the expectation values of observables by re-weighting although as will be seen in the next section the measured Pfaffian phase is always small in our simulations.

%%%%%%%%%%%%%%%%%%%%%%%%%%%%%%%%%%%%%%%%%%%%%%%%%%%%%%%%%%%
\subsection{Phase of the Pfaffian}
\label{sec:pfaf} 
%%%%%%%%%%%%%%%%%%%%%%%%%%%%%%%%%%%%%%%%%%%%%%%%%%%%%%%%%%%

The phase of the Pfaffian was studied in Ref. \cite{Kanamori:2007ye} for two different lattice constructions. Soon after, the phase of Pfaffian for the construction we use here was calculated in Ref. \cite{Catterall:2011aa} and it was observed that it vanishes as one approaches the continuum limit. It was correctly noted in Ref. \cite{Hanada:2010qg} that the absence of the sign is a property of the correct continuum limit. In this paper, we will study the phase of the Pfaffian at stronger couplings than have been explored before and on much larger lattices using the parallel code developed in Ref. \cite{Schaich:2014pda}. We show that the phase fluctuations become small and vanish as we take the continuum limit. This is true for all couplings we have considered. However, on a fixed lattice volume, the magnitude of the phase fluctuations grows with the coupling. This implies that accessing stronger couplings ($t \le 1/9$) requires the use of larger lattices if we are to avoid a sign problem. We show these results in Fig. \ref{fig:pfaffian1}. 
\subsection{Ground State Energy}
\label{sec:ground-state-e}
%%%%%%%%%%%%%%%%%%%%%%%%%%%%%%%%%%%%%%%%%%%%%%%%%%%%%%%%%%%

We now present our simulation results on the ground state energy of the theory. We would like to extrapolate the lattice data for ground state energy density $\frac{\vac}{N^2 \lambda}$ to zero temperature after taking the continuum ($a \to 0$) and massless ($\mu \to 0$) limits. A representative example of the mass extrapolations and continuum extrapolations are shown in Fig. \ref{fig:mass_U3} and Fig. \ref{fig:cont_U3}, respectively.  At the end, we perform three types of extrapolations in temperature - using power law, exponential, and constant fits. 

\begin{figure}
\begin{center} 
\includegraphics[width=0.45\textwidth]{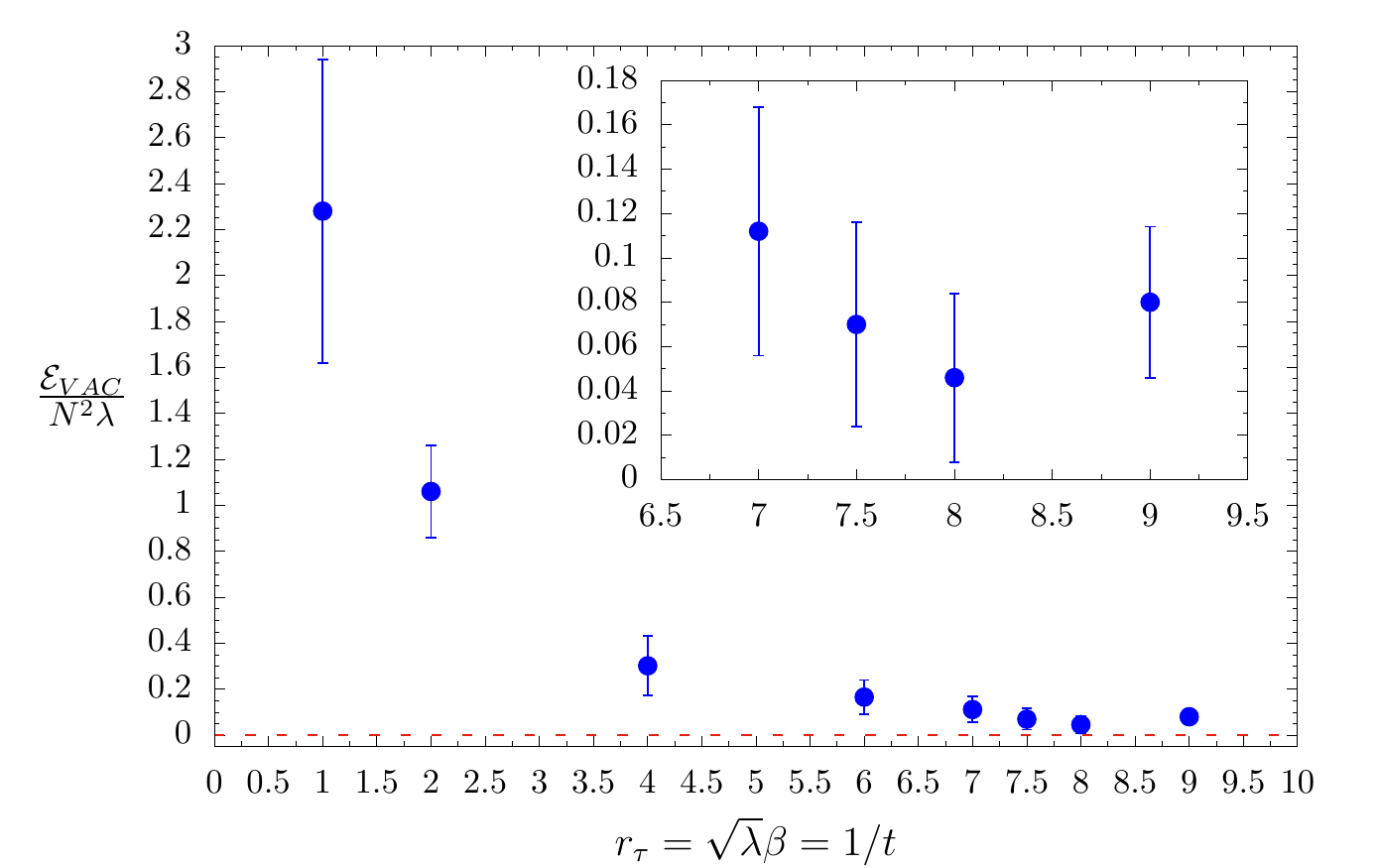}
\end{center}
\caption{\label{fig:beta_U2}The $\beta \to \infty$ extrapolation of the ground state energy for $U(2)$ gauge group. The inset zooms in to show the low-temperature regime. For details, see Table (\ref{tab:actionU2}).}
\end{figure}

\begin{figure}
\begin{center} 
\includegraphics[width=0.45\textwidth]{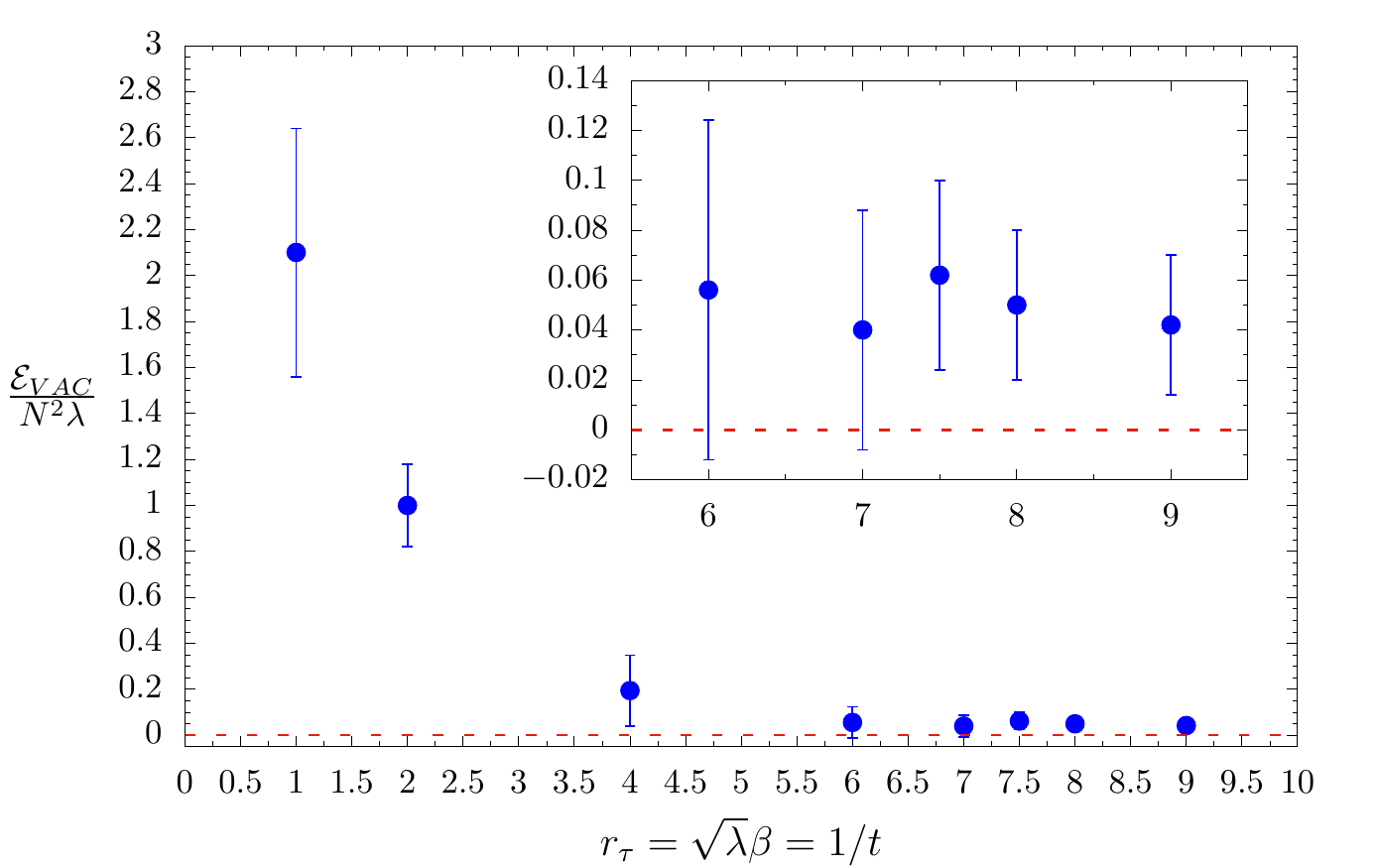}
\end{center}
\caption{\label{fig:beta_U3}The $\beta \to \infty$ extrapolation of the ground state energy for $U(3)$ gauge group. The inset zooms in to show the low-temperature regime. For details, see Table (\ref{tab:actionU3}).}
\end{figure}

We show the vacuum energy density vs inverse temperature for $U(2)$ in Fig.~\ref{fig:beta_U2}.
Extrapolating $\RT\to\infty$ using the range $\RT \in [6, 9]$
\beq
\frac{\vac}{N^2 \lambda} = \left\{
  \begin{array}{ll}
    0.06(4), ~ ~ \CHI = 0.40 & : \text{power law fit} \\      
    0.06(2), ~ ~ \CHI = 1.26 & : \text{exponential fit}\\ 
   0.08(2), ~ ~ \CHI = 0.63 & : \text{constant fit}\\  
  \end{array}
\right.
\eeq
In Fig.~\ref{fig:beta_U3} we show the vacuum energy density vs inverse temperature for gauge group $U(3)$.
Extrapolating $\RT\to\infty$ using the range $\RT \in [6, 9]$
\beq
\frac{\vac}{N^2 \lambda} = \left\{
  \begin{array}{ll}
    0.05(2), ~ ~ \CHI = 0.11 & : \text{power law fit} \\  
    0.04(4), ~ ~ \CHI = 0.11 & : \text{exponential fit} \\ 
    0.05(2), ~ ~ \CHI = 0.06 & : \text{constant fit} \\ 
  \end{array}
\right.
\eeq

%%%%%%%%%%%%%%%%%%%%%%%%%%%%%%%%%%%
%%%%%%%%%%%%%%%%%%%%%%%%%%%%%%%%%%%
%%%%%%%%%%%%%%%%%%%%%%%%%%%%%%%%%%%%%
%%%%%%%%%%%%%%%%%%%%%%%%%%%%%%%%%%%%%
We note that the errors in our results do not allow us to make conclusive statements about the exact form of the energy dependence on the temperature. Both power, exponential and constant fitting functions yield
comparable results consistent with vanishing ground state energy.  Our
calculation puts an upper bound on the dimensionless energy density using the constant fit at $\frac{\vac}{N^2 \lambda}=0.08(2)$ for U(2) and $\frac{\vac}{N^2 \lambda}=0.05(2)$ for U(3).

While this work was in progress results were presented on the tree-level $\cO(a)$ improvement of the Sugino's lattice action for two-dimensional $\cN = (2, 2)$ SYM \cite{Hanada:2017gqc}.
We note that our lattice formulation already possesses this improvement which we see in Fig.~\ref{fig:cont_U3} and in
Table \ref{table:order_a}.
%%%%%%%%%%%%%%%%%%%%%
\begin{table}[htbp]
   \vspace{10mm}
      \centering
 \begin{tabular}{cccccc}
\hline \hline
\ \ $\zeta$ \ \ & \ \ $\propto (a/L)^{p} $ \ \ & \ $\propto (a/L)^{p} + c $ \\
\hline
0.40 & 1.86(9) & 1.76(22) & \ \ \\
0.50 & 1.76(6) & 1.60(15) & \ \ \\
0.55 & 1.79(5) & 1.90(11) &  \ \ \\
0.60 & 1.74(4) & 1.70(11) &   \ \ \\
\hline
  \vspace{4mm}
\end{tabular}
      \centering
 \begin{tabular}{cccccc}

\hline \hline
\ \ $\zeta$ \ \ & \ \ $\propto (a/L)^{p} $ \ \ & \ $\propto (a/L)^{p} + c $ \\
\hline
0.40 & 1.73(10) & 1.58(24) & \ \ \\
0.50 & 1.71(7) & 1.74(17) & \\
0.55 & 1.69(6) & 1.57(14) &  \ \ \\
0.60 & 1.78(5) & 1.98(12) &   \ \ \\
\hline
\end{tabular}

\caption{\label{table:order_a}Numerical results showing that our action is effectively $\cO(a)$ improved. We measure the deviation of the bosonic action/site from its supersymmetric value of $\frac{3}{2}N^2$ and fit it to power law. The first column shows the soft-mass parameter, $\zeta$, we use to regulate the flat directions. The second column is the obtained value of the power, $p$, constraining vanishing intercept, the third is the obtained value of the power, $p$, \emph{without} constraining the intercept. We quote results from one of the couplings used in this work, $\RT$ = 6. On the top, we show the results with $U(3)$ and with $U(2)$ at the bottom. The fits are very good with maximum \CHI = 2.80.}
\end{table}
%%%%%%%%%%%%%%%%%%%%%

%%%%%%%%%%%%%%%%%%%%%%%%%%%%%%%%%%%%%%%%%%%%%%%%%%%%%%%%%%%
\section{Conclusions}
\label{sec:conclude}
%%%%%%%%%%%%%%%%%%%%%%%%%%%%%%%%%%%%%%%%%%%%%%%%%%%%%%%%%%%

In this paper we have examined the possibility of dynamical supersymmetry breaking in two-dimensional $\cN = (2, 2)$ SYM through lattice simulations. The lattice theory is exact supersymmetric, gauge invariant, local, and doubler free. 
We find an upper bound on the vacuum energy density of $\frac{\vac}{N^2 \lambda}=0.08(2)$ and $\frac{\vac}{N^2 \lambda}=0.05(2)$ for U(2) and U(3) respectively. The energy
density is statistically consistent with zero and hence with the absence of dynamical supersymmetry
breaking. It would be interesting to examine the spectrum in future work to confirm the absence of spontaneous supersymmetry
breaking perhaps by searching for signals of a Goldstino as was done in \cite{Catterall:2015tta}. We have also measured the phase of the Pfaffian on all our ensembles and find that while the average
phase grows with coupling it decreases as we take the continuum limit in agreement with theoretical expectations.
In practice, it is numerically small for all our ensembles. 
The question of supersymmetry breaking in this model was addressed before in \cite{Kanamori:2009dk}. Our current work, in addition to using a different lattice action, has employed stronger couplings (and hence
lower temperatures) and much smaller lattice spacings. For example, the lowest temperature used in the earlier work was $t = 1/6$ as compared to $t = 1/9$ in this work while the largest lattice used here is $96\times 96$ as compared
to $30\times 12$ in the earlier study.

%%%%%%%%%%%%%%%%%%%%%%%%%%%%%%%%%%%%%%%%%%%%%%%%%%%%%%%%%%%
\section*{ACKNOWLEDGEMENTS} 
%%%%%%%%%%%%%%%%%%%%%%%%%%%%%%%%%%%%%%%%%%%%%%%%%%%%%%%%%%%

SC and RGJ were supported by the US Department of Energy (DOE), Office of Science, Office of High Energy Physics, under Award Number DE-SC0009998. AJ gratefully acknowledges support from the International Centre for Theoretical Sciences (ICTS-TIFR), the Infosys Foundation and the Indo-French Centre for the Promotion of Advanced Research (IFCPAR/CEFIPRA). Numerical calculations were carried out on the DOE-funded USQCD facilities at Fermilab, and on the Mowgli cluster at ICTS-TIFR. AJ thanks Srinivasa R. and Mohammad Irshad for providing timely technical support with the usage of the computing facility at ICTS-TIFR. RGJ would like to thank Issaku Kanamori for discussions during a workshop at YITP, Kyoto in April 2017.  

%%%%%%%%%%%%%
\bibliographystyle{utphys}
\bibliography{paper}
%%%%%%%%%%%%%

\appendix

\begin{table*}[ht]
\section{Data Tables for $U(2)$ and $U(3)$}
\centering
\resizebox{1.5\columnwidth}{!}{
\renewcommand\arraystretch{1.0}   
  \addtolength{\tabcolsep}{.4 pt}  
\begin{tabular}{c@{\hspace{.45em}}c@{\hspace{1.1em}}c@{\hspace{1.1em}}c@{\hspace{-.5em}}c@{\hspace{-.5em}}cc}
\hline\hline
 $\RT$ & $N_{x} \times N_{t}$ & $\left. -\SB/N^2 \lam \right\vert_{ \zeta=0.4 }$ & $\left. -\SB/N^2 \lam \right\vert_{ \zeta=0.5 }$ & $\left. -\SB/N^2 \lam \right\vert_{ \zeta=0.55 }$  & $\left. -\SB/N^2 \lam \right\vert_{ \zeta=0.6 }$  &  \\
\hline \hline
1.0 & $ 24 \times 24$  & 1.14(33)  & --- & ---  & --- \\
\hline  \hline  
\vspace{10mm} 
2.0 & $ 24 \times 24$  & 0.53(10)  & ---  & ---  & --- \\
\hline  \hline 
4.0 & $ 24 \times 24$  & 0.253(21)  & 0.332(23) & 0.440(20)  & 0.502(21) \\
& $ 32 \times 32$  & 0.272(34)  & 0.355(35)  &  0.407(35) & 0.501(33)  \\
& $ 48 \times 48$ & 0.354(43)   &  0.378(48) & 0.531(45)  & 0.538(45)  \\
& $ 96 \times 96$ & 0.26(10)   &  0.40(10) & 0.48(10)  & 0.63(10)  \\
& \cline{4-6} &  & & & $ \lim_{\mathrm{a} \to 0}$ , then $\lim_{\mu \to 0} = 0.151(65)$  \\
&  & & & $\lim_{\mu \to 0}$  , then $\lim_{\mathrm{a} \to 0} =  0.148(65)$ \\
\hline  \hline
6.0 & $ 24 \times 24$  & 0.20(1)  & 0.30(1) & 0.35(1) & 0.43(1) \\
& $ 32 \times 32$  & 0.22(1)  & 0.32(2) & 0.39(2) & 0.44(1) \\
& $ 48 \times 48$ & 0.25(2)  & 0.37(2) & 0.44(2) & 0.48(2) \\
& $ 96 \times 96$ & 0.27(5)  & 0.45(5) & 0.48(2) & 0.62(5) \\
& \cline{4-6} &  & & & $\lim_{\mathrm{a} \to 0}$, then $\lim_{\mu \to 0} = 0.083(37) $  \\
&  & & & $\lim_{\mu \to 0}$  , then $\lim_{\mathrm{a} \to 0} =  0.079(38) $ \\
\hline  \hline 
7.0 & $ 24 \times 24$  & 0.20(1)  & 0.28(1) & 0.33(1)  & 0.38(1) \\
& $ 32 \times 32$  & 0.22(1)  & 0.30(1) & 0.38(1)  & 0.43(1)  \\
& $ 48 \times 48$ & 0.25(1)  & 0.34(2) & 0.41(2)  & 0.47(2)  \\
 & $ 96 \times 96$ & 0.27(4)  & 0.38(3)  & 0.45(3)  &  0.56(4) \\
& \cline{4-6} &  & & & $\lim_{\mathrm{a} \to 0}$, then $\lim_{\mu \to 0} = 0.056(28) $  \\
&  & & & $\lim_{\mu \to 0}$  , then $\lim_{\mathrm{a} \to 0} =  0.055(28) $ \\
\hline  \hline
7.5 & $ 24 \times 24$  & 0.19(1)  & 0.29(1) & 0.33(5)  &  0.38(1) \\
& $ 32 \times 32$  &  0.19(1)  & 0.31(1) & 0.36(1)  & 0.42(1) \\
& $ 48 \times 48$ &  0.23(1) & 0.34(2) & 0.39(1) & 0.47(2)  \\
 & $ 96 \times 96$ & 0.24(3)  & 0.40(3) &  0.45(3) & 0.48(3)  \\
& \cline{4-6} &  & & & $\lim_{\mathrm{a} \to 0}$, then $\lim_{\mu \to 0} = 0.035(23) $  \\
&  & & & $\lim_{\mu \to 0}$  , then $\lim_{\mathrm{a} \to 0} =  0.033(24) $ \\
\hline  \hline
8.0 & $ 24 \times 24$  & 0.20(1)  & 0.28(1)  & 0.33(1)  & 0.36(1) \\
& $ 32 \times 32$  &  0.20(1) & 0.29(1) & 0.35(1)  &  0.40(1) \\
& $ 48 \times 48$ &  0.22(1) & 0.34(1) & 0.37(1)  & 0.44(1) \\
 & $ 96 \times 96$ &  0.27(2) & 0.36(3) & 0.44(3)   & 0.54(3) \\
& \cline{4-6} &  & & & $\lim_{\mathrm{a} \to 0}$, then $\lim_{\mu \to 0} = 0.023(19) $  \\
&  & & & $\lim_{\mu \to 0}$  , then $\lim_{\mathrm{a} \to 0} =  0.022(20) $ \\
\hline  \hline
9.0 & $ 24 \times 24$  & 0.184(5)  & 0.260(4) &  0.304(4) & 0.361(4) \\
& $ 32 \times 32$  & 0.20(1)  & 0.30(1) & 0.35(1)  & 0.39(1) \\
& $ 48 \times 48$ &  0.22(1) & 0.32(1) &  0.36(1) & 0.45(1) \\
 & $ 96 \times 96$ &  0.254(22) & 0.364(17) & 0.435(20)  & 0.534(22) \\
& \cline{4-6} &  & & & $\lim_{\mathrm{a} \to 0}$, then $\lim_{\mu \to 0} = 0.040(17) $  \\
&  & & & $\lim_{\mu \to 0}$  , then $\lim_{\mathrm{a} \to 0} =  0.039(17) $ \\
\hline  \hline
\end{tabular}
}
\caption{
\label{tab:actionU2}The action density, $-\SB/N^2 \lambda$, which is related to the ground state energy density using Eq. (\ref{eq:energy1}) or Eq. (\ref{eq:energy2}) with gauge group $U(2)$ for different lattices, mass parameters and couplings used in this work. The results are obtained through blocked jackknife analyses. We have considered at least 5000 thermalized molecular dynamics time units (MDTU) in each case.} 
\end{table*}
%%%%%%%%%%%%%%%%%%%%%%%%%%%%%%

%%%%%%%%%%%%%%%%%%%%%%%%%%%%%%
\begin{table*}[ht]
\centering
\resizebox{1.5\columnwidth}{!}{
\renewcommand\arraystretch{1.0}   
  \addtolength{\tabcolsep}{1 pt}  
\begin{tabular}{c@{\hspace{.45em}}c@{\hspace{1.1em}}c@{\hspace{1.1em}}c@{\hspace{-.5em}}c@{\hspace{-.5em}}cc}
\hline\hline
 $\RT$ & $N_{x} \times N_{t}$ & $\left. -\SB/N^2 \lam \right\vert_{ \zeta=0.4 }$ & $\left. -\SB/N^2 \lam \right\vert_{ \zeta=0.5 }$ & $\left. -\SB/N^2 \lam \right\vert_{ \zeta=0.55 }$  & $\left. -\SB/N^2 \lam \right\vert_{ \zeta=0.6 }$  &  \\
\hline \hline
1.0 & $ 24 \times 24$  & 1.05(27)  & --- & ---  & --- \\
\hline  \hline  
\vspace{10mm} 
2.0 & $ 24 \times 24$  & 0.50(9)  & ---  & ---  & --- \\
\hline  \hline 
4.0 & $ 24 \times 24$  & 0.228(17)  & 0.329(18) & 0.405(19)  & 0.443(18) \\
& $ 32 \times 32$  & 0.235(26)  & 0.370(27)  &  0.445(22) & 0.512(25)  \\
& $ 48 \times 48$ & 0.302(41)   &  0.460(37) & 0.484(37)  & 0.531(32)  \\
 & $ 96 \times 96$ & 0.164(60)   & 0.423(78)  & 0.665(78)  & 0.631(78) \\
& \cline{4-6} &  & & & $ \lim_{\mathrm{a} \to 0}$ , then $\lim_{\mu \to 0} = 0.097(77) $  \\
&  & & & $\lim_{\mu \to 0}$  , then $\lim_{\mathrm{a} \to 0} =  0.074(54) $ \\
\hline  \hline
6.0 & $ 24 \times 24$  & 0.205(8)  & 0.297(10) & 0.363(8)  & 0.412(8) \\
& $ 32 \times 32$  & 0.228(11)  & 0.332(10)  &  0.391(11) & 0.428(12)  \\
& $ 48 \times 48$ & 0.220(24)   &  0.366(21) & 0.397(17)  & 0.515(20)  \\
 & $ 96 \times 96$ & 0.246(35)   & 0.399(35)  & 0.511(35)  & 0.580(31) \\
& \cline{4-6} &  & & & $\lim_{\mathrm{a} \to 0}$, then  $\lim_{\mu \to 0} = 0.028(34) $  \\
&  & & & $\lim_{\mu \to 0}$  , then $\lim_{\mathrm{a} \to 0} =  0.034(34) $ \\
\hline  \hline 
7.0 & $ 24 \times 24$  &  0.206(7) & 0.285(5) &  0.335(5) & 0.394(6) \\
& $ 32 \times 32$  & 0.195(9)  &0.309(9)  & 0.351(10)  & 0.406(11) \\
& $ 48 \times 48$ & 0.231(13)  & 0.327(13) & 0.387(13)  &0.473(11)  \\
 & $ 96 \times 96$ & 0.262(28)  &0.463(34)  &  0.440(25) & 0.550(28) \\
& \cline{4-6} &  & & & $\lim_{\mathrm{a} \to 0}$, then $\lim_{\mu \to 0} = 0.020(24) $  \\
&  & & & $\lim_{\mu \to 0}$  , then $\lim_{\mathrm{a} \to 0} =  0.026(24) $ \\
\hline  \hline 
7.5 & $ 24 \times 24$  &  0.196(5) & 0.275(5) & 0.330(5)  & 0.376(5) \\
& $ 32 \times 32$  &0.214(7)   & 0.299(9) & 0.358(8)  & 0.418(9) \\
& $ 48 \times 48$ & 0.23(1)  & 0.33(1) & 0.40(1)   & 0.46(1)  \\
 & $ 96 \times 96$ &  0.24(2) & 0.39(2) &  0.49(2) & 0.55(3) \\
& \cline{4-6} &  & & & $\lim_{\mathrm{a} \to 0}$, then $\lim_{\mu \to 0} = 0.031(19) $  \\
&  & & & $\lim_{\mu \to 0}$  , then $\lim_{\mathrm{a} \to 0} =  0.030(20) $ \\
\hline  \hline 
8.0 & $ 24 \times 24$  & 0.192(5)  & 0.273(5) &  0.324(5) & 0.367(3)  \\
& $ 32 \times 32$  & 0.210(6)  & 0.300(6) & 0.347(6)  & 0.413(7) \\
& $ 48 \times 48$ & 0.212(9)  &0.340(9)  &  0.408(13) & 0.463(7)  \\
 & $ 96 \times 96$ & 0.261(15)  & 0.387(19) & 0.471(17)  & 0.529(15) \\
& \cline{4-6} &  & & & $\lim_{\mathrm{a} \to 0}$, then $\lim_{\mu \to 0} = 0.025(15) $  \\
&  & & & $\lim_{\mu \to 0}$  , then $\lim_{\mathrm{a} \to 0} =  0.026(16) $ \\
\hline  \hline 
9.0 & $ 24 \times 24$  & 0.182(4)  & 0.259(4) & 0.312(3)  & 0.351(4)  \\
& $ 32 \times 32$  & 0.199(6)  & 0.288(6) & 0.331(6)  & 0.394(5) \\
& $ 48 \times 48$ &0.207(7)   & 0.314(7) &0.380(7)   &0.446(9)  \\
 & $ 96 \times 96$ & 0.246(17)  &  0.358(12)& 0.444(17)  &0.521(17)  \\
& \cline{4-6} &  & & & $\lim_{\mathrm{a} \to 0}$, then $\lim_{\mu \to 0} = 0.021(14) $  \\
&  & & & $\lim_{\mu \to 0}$  , then $\lim_{\mathrm{a} \to 0} =  0.016(14) $ \\
\hline  \hline 
\end{tabular}
}
\caption{
The action density, $-\SB/N^2 \lambda$, which is related to the ground state energy density using Eq. (\ref{eq:energy1}) or Eq. (\ref{eq:energy2}) with gauge group $U(3)$ for different lattices, mass parameters and couplings used in this work. The results are obtained through blocked jackknife analyses. We have considered at least 4000 thermalized molecular dynamics time units (MDTU) in each case.} 
\label{tab:actionU3}
\end{table*}

%%%%%%%%%%%%%%%%%%%%%%%%%%%%%%%%%%%%%%%%%%%%

\end{document}